\documentclass[sn-mathphys,Numbered]{sn-jnl}

\usepackage{graphicx}%
\usepackage{multirow}%
\usepackage{amsmath,amssymb,amsfonts}%
\usepackage{amsthm}%
\usepackage{mathrsfs}%
\usepackage[title]{appendix}%
\usepackage{xcolor}%
\usepackage{textcomp}%
\usepackage{manyfoot}%
\usepackage{booktabs}%
\usepackage{algorithm}%
\usepackage{algorithmicx}%
\usepackage{algpseudocode}%
\usepackage{listings}%

\theoremstyle{thmstyleone}%
\theoremstyle{thmstyletwo}%

\theoremstyle{thmstylethree}%
\newcommand{\RNum}[1]{\uppercase\expandafter{\romannumeral #1\relax}}    
\begin{document}

\title[Seebeck power generation and Peltier cooling in a Normal metal-quantum dot-superconductor nanodevice]{Seebeck power generation and Peltier cooling in a Normal metal-quantum dot-superconductor nanodevice}


\author*[1]{\fnm{Sachin} \sur{Verma}}\email{sverma2@ph.iitr.ac.in}

\author*[1]{\fnm{Ajay} \sur{}}\email{ajay@ph.iitr.ac.in}

\affil[1]{\orgdiv{Department of Physics}, \orgname{India Institute of Technology Roorkee}, \postcode{247667}, \state{Uttarakhand}, \country{India}}


\abstract{We theoretically investigate the Seebeck and Peltier effect across an interacting quantum dot(QD) coupled between a normal metal and a Bardeen-Cooper-Schrieffer superconductor within the Coulomb blockade regime. Our results demonstrate that the thermoelectric conversion efficiency at optimal power output (optimized with respect to QD energy level and external serial load) in NQDS nanodevice can reach up to \unboldmath$58\%\eta_C$, where \unboldmath$\eta_C$ is Carnot efficiency, with output power \unboldmath$P_{max}\approx 35fW$ for temperature below the superconducting transition temperature. Further, the Peltier cooling effect is observed for a wide range of parameter regimes, which can be optimized by varying the background thermal energy, QD level energy, QD-reservoir tunneling strength, and bias voltage. The results presented in this study are within the scope of existing experimental capabilities for designing miniature hybrid devices that operate at cryogenic temperatures.}




\maketitle

\section{Introduction}
\label{sec:1}
In recent years thermoelectric effects and heat transport in superconductor-quantum dot(QD) based hybrid nanodevices has received significant attention due to its potential applications in energy harvesting and cooling at the nanoscale\cite{Krawiec2008, Wysokinski2012, Hwang2015, Sothmann2015, Hwang2016a, Hwang2016b, Sanchez2016, Xu2016, Kleeorin2016,Nie2016, Benenti2017, Hwang2017, Barnas2017,  Hwang2018,Yao2018a, Yao2018b, Kamp2019, Verma2022, Tabatabaei2022, Yao2023, Kumar2023, Hwang2023}. These hybrid devices combine quantum dots, superconductors, normal metals, or ferromagnetic materials, and have emerged as promising platforms for studying charge and heat transport at the nanoscale. The addition of superconducting components introduces novel features, such as diverging quasiparticle density of state near superconducting energy gap edge, the formation of subgap Andreev bound states, and proximity-induced superconductivity, which can significantly impact the transport properties of the system. The operation of these hybrid mesoscopic devices is based on phenomena that appear only at cryogenic temperatures. Thus, these devices must be cooled down to a few Kelvin or lower temperatures. However, it also has the advantage that these low-temperature devices are often more efficient than their bulk counterparts. Further, the gate-tunable discrete energy levels of QD serve as perfect energy filters for electron transport, resulting in improved thermoelectric performance.\\
In superconductor-QD nanodevices, the Seebeck effect, which describes the generation of a voltage difference across a temperature gradient, i.e., system works as a particle exchange heat engine or power generator, has been studied in a few papers\cite{Krawiec2008,Hwang2016b,Barnas2017,Verma2022,Tabatabaei2022}. Recently, hybrid superconductor-QD nanodevice have also been found to exhibit the Peltier effect, which refers to the generation or absorption of heat at the QD-reservoirs junction, and system works as a cooler\cite{Hwang2023}. Few studies have also explored the charge and spin Seebeck diode effect\cite{Hwang2016a}, cross thermoelectric effect\cite{Hwang2015}, heat transport\cite{Kamp2019,Hwang2018,Verma2022}, and thermophase Seebeck effect\cite{Kleeorin2016,Kumar2023} in hybrid superconductor-QD nanodevices. Moreover, thermoelectric effects and heat transport in multi-terminal and multi-dot configurations in the presence of superconducting component have also been explored by several authors\cite{Wysokinski2012, Xu2016,Nie2016,Yao2018a,Yao2018b,Yao2023}.\\
In the present work, we examine non-linear thermoelectric transport in NQDS nanodevices using the equation of motion technique within the Hubbard-\RNum{1} approximation and Keldysh non-equilibrium Green's function (NEGF) formalism. The optimal performance of the superconductor-QD-based thermoelectric particle exchange heat engine has not been studied so far. Therefore, first, we examined the optimal performance of an interacting normal metal-quantum dot-superconductor (NQDS) hybrid heat engine in the presence of an external serial load. Secondly, we study the heat current and Peltier cooling effect in NQDS nanodevice as a function of different system parameters.\\
This paper is structured as follows: The preceding section \hyperref[sec:2]{2} discusses the effective model Hamiltonian and theoretical formalism. Section \hyperref[sec:3]{3} contains the numerical results and explanation for heat and thermoelectric transport. Section \hyperref[sec:4]{4} concludes the present work.
\section{Model Hamiltonian and theoretical description}
\label{sec:2}
The NQDS system is modelled by single impurity Anderson model and Bogoliubov transformed BCS mean-field Hamiltonian\cite{Verma2022},
\begin{equation} \label{eq:1}
\begin{aligned}
H = & \sum_{k,\sigma}\epsilon_{k,N}c^\dagger_{k\sigma,N}c_{k\sigma,N}+\sum_{k,\sigma}E_{k}\gamma^\dagger_{k\sigma}\gamma_{k\sigma}+\sum_{\sigma} \epsilon_{d}n_{\sigma}+Un_{\uparrow}n_{\downarrow}+\\[4pt]
&\sum_{k\sigma}(\mathcal{V}_{k,N}d^\dagger_{\sigma}c_{k\sigma,N}+\mathcal{V}^{\ast}_{k,N}c^\dagger_{k\sigma,N}d_{\sigma})+\sum_{k\sigma}(\mathcal{V}_{k,S}u^\ast_k d^\dagger_{\sigma}\gamma_{k\sigma}+\mathcal{V}^{\ast}_{k,S}u_k\gamma^{\dagger}_{k\sigma}d_{\sigma})+\\[4pt]
&\sum_{k}[\mathcal{V}^\ast_{k,S}v_k(d^\dagger_{\uparrow}\gamma^\dagger_{-k\downarrow}-d^\dagger_{\downarrow}\gamma^\dagger_{k\uparrow})+\mathcal{V}_{k,S}v^{\ast}_k(\gamma_{-k\downarrow}d_{\uparrow}-\gamma_{k\uparrow}d_{\downarrow})]
\end{aligned}
\end{equation}
The first term describes the normal metallic  reservoir in the non-interacting quasi-particle approximation with   single electron kinetic energy $\epsilon_{k,N}$ and $c_{k\sigma,N}(c^\dagger_{k\sigma,N})$ is the annihilation(creation) operator of an electron with spin $\sigma$ and wave vector $\vec{k}$.\\
The second term describes the superconducting reservoir, where $\gamma_{k\sigma}(\gamma^\dagger_{k\sigma})$ is the annihilation(creation) operator for Bogoliubov quasiparticles with spin $\sigma$, wave vector $\vec{k}$ and energy $E_{k} = \sqrt{\epsilon^2_{k,S}+|\Delta|^2}$. The temperature dependence of the superconducting energy gap is given as $\Delta(T)=\Delta_0\,tanh\left(1.74\sqrt{(T_c/T)-1}\right)$, where $\Delta_0$ is superconducting energy gap at absolute zero temperature and $T_c$ is critical temperature with $k_BT_c=0.568\Delta_0$.\\
Third term  describes the Hamiltonian for single-level QD with energy $\epsilon_d$, and $d_\sigma(d^\dagger_\sigma)$ is the annihilation(creation) operator of electron with spin $\sigma$ on the QD and $n_{\sigma}=d_\sigma^\dagger d_\sigma$ is number operator. The QD can have maximum occupancy of two electrons with opposite spins. We also consider the intradot electron-electron Coulomb repulsion with the interaction strength $U$ represented by the fourth term.\\
The remaining terms represents the tunnelling Hamiltonian between the QD energy level and reservoirs with $\mathcal{V}_{k\alpha}$ as the tunnelling amplitude between the QD and the $\alpha$-reservoir ($\alpha\in N,S$). The coefficients $u_k$ and $v_k$ read
\begin{equation}  \label{eq:2}
|{u_k}|^2=\frac{1}{2}\left(1+\frac{\epsilon_{k,S}}{\sqrt{{\epsilon^2_{k,S}}+|\Delta|^2}}\right)\quad\&\quad |{v_k}|^2=\frac{1}{2}\left(1-\frac{\epsilon_{k,S}}{\sqrt{{\epsilon^2_{k,S}}+|\Delta|^2}}\right)
\end{equation}
In order to study the thermoelectric transport properties of NQDS system using model Hamiltonian in Eqn. \eqref{eq:1}, we apply Green’s function equation of motion technique. To truncate hierarchy of Green’s function equation of motions, we use Hubbard-\RNum{1} approximation\cite{Hubbard1963,Verma2022}, which is good enough to describe the Coulomb blockade regime at temperatures $T>>T_K$, where $T_K$ is Kondo temperature. Furthermore, we assume that the coupling strength $\mathcal{V}_{k,\alpha}$ is $k$ independent and is much smaller than the half bandwidth $D\rightarrow\infty$. Therefore, the tunneling rate from the dot to the $\alpha$-leads is represented as $\Gamma_{\alpha}=2\pi|\mathcal{V}_{\alpha}|^2\rho_{0\alpha}$, where the density of states in normal metallic state, denoted by $\rho_{0\alpha}$, remains constant within a range of energy around the Fermi level.\\
In Nambu representation, we define the single particle retarded Green's function of the QD as a $2\times 2$ matrices
 \begin{equation} \label{eq:3}
 {\bf{G}}^{r}_{d}(\omega)={\left\langle\left\langle{
\begin{pmatrix}
d_{\uparrow}\\
d_{\downarrow}^{\dagger}\\
\end{pmatrix}
\begin{pmatrix}
d_{\uparrow}^{\dagger} & d_{\downarrow}
\end{pmatrix}
}\right\rangle\right\rangle}=
\begin{pmatrix}
 \langle\langle{d_{\uparrow}|d_{\uparrow}^{\dagger}}\rangle\rangle & \langle\langle{d_{\uparrow}|d_{\downarrow}}\rangle\rangle \\
 \langle\langle{d_{\downarrow}^{\dagger}|d_{\uparrow}^{\dagger}}\rangle\rangle & \langle\langle{d_{\downarrow}^{\dagger}|d_{\downarrow}}\rangle\rangle \\
\end{pmatrix}=
\begin{pmatrix}
 G^r_{d,11}(\omega) & G^r_{d,12}(\omega) \\
 G^r_{d,21}(\omega) & G^r_{d,22}(\omega) \\
\end{pmatrix}
\end{equation}
Where the diagonal components of ${\bf{G}}^{r}_{d}(\omega)$ represents the single particle retarded Green's function of electron with spin $\sigma=\uparrow$ and hole with spin $\sigma=\downarrow$  respectively. The off-diagonal component represents the superconducting paring correlation on the QD. The Fourier transform of the single particle retarded Green's function for QD
\begin{equation*}
\begin{aligned}
G_{d,11}^r(\omega)=\langle\langle{d_{\uparrow}|d^\dagger_{\uparrow}}\rangle\rangle\nonumber=-\cfrac{i}{2\pi}\lim_{\delta \to 0^{+}}\int{\theta(t)\langle\{d_{\uparrow}(t),d^\dagger_{\uparrow}(0)\}\rangle e^{i(\omega+i\delta)t}}dt
\end{aligned}
\end{equation*}
where $\theta(t)$ is Heaviside function, must satisfies the following EOM
\begin{equation} \label{eq:4}
\omega\langle\langle{d_{\uparrow}|d^\dagger_{\uparrow}}\rangle\rangle=\langle\{d_{\sigma},d^\dagger_{\uparrow}\}\rangle+\langle\langle[d_{\uparrow},H]|d^\dagger_{\uparrow}\rangle\rangle.
\end{equation}
By evaluating different commutator and anti-commutator brackets we drive the following EOMs for the single particle Green's functions
\begin{multline}\label{eq:5}
\left\{\omega-\epsilon_d-\sum_k\frac{|\mathcal{V}_{k,N}|^2}{\omega-\epsilon_{k,N}}-\sum_k|\mathcal{V}_{k,S}|^2\left(\frac{|u_k|^2}{\omega-E_k}+\frac{|v_k|^2}{\omega+E_k}\right)\right\}\langle\langle{d_{\uparrow}|d_{\uparrow}^{\dagger}}\rangle\rangle = \\[8pt]
1+\left\{{\sum_k|\mathcal{V}_{k,S}|^2u^\ast_kv_k\left(\frac{1}{\omega-E_k}-\frac{1}{\omega+E_k}\right)}\right\}\langle\langle{d_{\downarrow}^{\dagger}|d_{\uparrow}^{\dagger}}\rangle\rangle+U\langle\langle{d_{\uparrow}n_{\downarrow}|d_{\uparrow}^{\dagger}}\rangle\rangle
\end{multline}
\begin{multline}\label{eq:6}
\left\{\omega+\epsilon_d-\sum_k\frac{|\mathcal{V}_{k,N}|^2}{\omega+\epsilon_{k,N}}-\sum_k|\mathcal{V}_{k,S}|^2\left(\frac{|u_k|^2}{\omega+E_k}+\frac{|v_k|^2}{\omega-E_k}\right)\right\}\langle\langle{d_{\downarrow}^{\dagger}|d_{\uparrow}^{\dagger}}\rangle\rangle = \\[8pt]
\left\{{\sum_k|\mathcal{V}_{k,S}|^2u_kv^\ast_k\left(\frac{1}{\omega-E_k}-\frac{1}{\omega+E_k}\right)}\right\}\langle\langle{d_{\uparrow}|d_{\uparrow}^{\dagger}}\rangle\rangle-U\langle\langle{d_{\downarrow}^{\dagger}n_{\uparrow}|d_{\uparrow}^{\dagger}}\rangle\rangle
\end{multline}
\begin{multline}\label{eq:7}
\left\{\frac{\omega-\epsilon_d-U}{\langle{n_{\downarrow}}\rangle}\right\}\langle\langle{d_{\uparrow}n_{\downarrow}|d_{\uparrow}^{\dagger}}\rangle\rangle = 1+\left\{{\sum_k|\mathcal{V}_{k,S}|^2u^\ast_kv_k\left(\frac{1}{\omega-E_k}-\frac{1}{\omega+E_k}\right)}\right\}\langle\langle{d_{\downarrow}^{\dagger}|d_{\uparrow}^{\dagger}}\rangle\rangle\\[8pt]
+\left\{\sum_k\frac{|\mathcal{V}_{k,N}|^2}{\omega-\epsilon_{k,N}}+\sum_k|\mathcal{V}_{k,S}|^2\left(\frac{|u_k|^2}{\omega-E_k}+\frac{|v_k|^2}{\omega+E_k}\right)\right\}\langle\langle{d_{\uparrow}|d_{\uparrow}^{\dagger}}\rangle\rangle
\end{multline}
\begin{multline}\label{eq:8}
\left\{\frac{\omega+\epsilon_d+U}{\langle{n_{\uparrow}}\rangle}\right\}\langle\langle{d_{\downarrow}^{\dagger}n_{\uparrow}|d_{\uparrow}^{\dagger}}\rangle\rangle = \left\{{\sum_k|\mathcal{V}_{k,S}|^2u_kv^\ast_k\left(\frac{1}{\omega-E_k}-\frac{1}{\omega+E_k}\right)}\right\}\langle\langle{d_{\uparrow}|d_{\uparrow}^{\dagger}}\rangle\rangle\\[8pt]
+\left\{\sum_k\frac{|\mathcal{V}_{k,N}|^2}{\omega+\epsilon_{k,N}}+\sum_k|\mathcal{V}_{k,S}|^2\left(\frac{|u_k|^2}{\omega-E_k}+\frac{|v_k|^2}{\omega+E_k}\right)\right\}\langle\langle{d_{\downarrow}^{\dagger}|d_{\uparrow}^{\dagger}}\rangle\rangle
\end{multline}
The terms with summations over $k$ appearing in above equations can be simplified by replacing $\sum_k \rightarrow \int{\rho(\epsilon)d\epsilon}$ and then solving these expressions using the complex contour  integration in the flat wide band limit. Finally after solving coupled Eqns. \eqref{eq:5}-\eqref{eq:8} we arrive at the expression for the retarded Green's function of electron with spin $\sigma=\uparrow$ and off-diagonal superconducting pairing correlation on the QD i.e.,
\begin{equation}\label{eq:9}
G_{d,11}^{r}(\omega) =\cfrac{\alpha_1(\omega)}{\omega-\epsilon_d+\left(\cfrac{i\Gamma_N}{2}+\beta(\omega)\right)\alpha_1(\omega)-\cfrac{\alpha_1(\omega)\,\alpha_2(\omega)\left(\cfrac{\Delta}{|\omega|}\beta(\omega)\right)^2}{\omega+\epsilon_d+\left(\cfrac{i\Gamma_N}{2}+\beta(\omega)\right)\alpha_2(\omega)}}
\end{equation}\\
\begin{equation}\label{eq:10}
G_{d,21}^{r}(\omega) =\cfrac{\alpha_2(\omega)\left(\cfrac{\Delta}{|\omega|}\beta(\omega)\right)}{\omega+\epsilon_d+\left(\cfrac{i\Gamma_N}{2}+\beta(\omega)\right)\alpha_2(\omega)}\times G_{d,11}^{r}(\omega)
\end{equation}
where\\
$${\alpha_1(\omega)=1+\cfrac{U\langle{n_{\downarrow}}\rangle}{\omega-\epsilon_d-U}}\;,\quad{\alpha_2(\omega)=1+\cfrac{U\langle{n_{\uparrow}}\rangle}{\omega+\epsilon_d+U}}$$ and 
\begin{equation*}
\beta(\omega)=\cfrac{\Gamma_S}{2}\rho_S(\omega)=\cfrac{\Gamma_S}{2}\cfrac{\omega}{\sqrt{\Delta^2-\omega^2}}\theta(\Delta-|\omega|)+\cfrac{i\Gamma_S}{2}\cfrac{|\omega|}{\sqrt{\omega^2-\Delta^2}}\theta(|\omega|-\Delta)
\end{equation*}
with $\rho_S(\omega)$ as the modified BCS density of states. The other matrix elements is given by $G_{d,22}^{r}(\omega)=-G_{d,11}^{r}(-\omega)^{\ast}$ and $G_{d,12}^{r}(\omega)=G_{d,21}^{r}(-\omega)^{\ast}$. These retarded Green's functions allow us to calculate the advanced and lesser/greater Green's functions and eventually the thermoelectric transport properties.\\
The  average occupancy on the quantum dot ($\langle{n_{\uparrow}}\rangle$=$\langle{n_{\downarrow}}\rangle$ for non-magnetic system) is calculated using the self-consistent integral equation of the form
\begin{equation} \label{eq:11}
\langle{n_{\sigma}}\rangle=\frac{-i}{2\pi}\int^{\infty}_{-\infty}G^{<}_{d,11}(\omega) d\omega
\end{equation}
where the lesser Green's function $G^{<}_{d}$ obeys the Keldysh equation\cite{Keldysh1965,Haug2008}
 \begin{equation} \label{eq:12}
{\bf{G}}^{<}_{d\sigma}(\omega)={\bf{G}}^{r}_{d\sigma}(\omega){\bf{\Sigma}}^{<}(\omega){\bf{G}}^{a}_{d\sigma}(\omega).
\end{equation}
The advanced Green's function matrix is ${\bf{G}}^{a}_{d\sigma}(\omega)=\left[{\bf{G}}^{r}_{d\sigma}(\omega)\right]^{\dagger}$ and the lesser self energy matrix is obtained using Ng ansatz\cite{Ng1996,Haug2008} i.e. 
\begin{equation}\label{eq:13}
{\bf{\Sigma}}^{<}(\omega)= -\sum_{\alpha\in N,S}\left[{\bf{\Sigma}}^{r}_{\alpha}-{\bf{\Sigma}}^{a}_{\alpha}\right]f_{\alpha}(\omega-\mu_{\alpha})
\end{equation}
This ansatz satisfies the continuity equation in steady state, allowing us to derive the lesser Green's function to examine the transport properties.\\
Now using retarded and advanced self-energy we get
\begin{equation}\label{eq:14}
{\bf{\Sigma}}^{<}(\omega)=
\begin{pmatrix}
\;\Sigma^{<}_{11}(\omega) &\quad \Sigma^{<}_{12}(\omega)\;\\[10pt]
\;\Sigma^{<}_{21}(\omega) &\quad \Sigma^{<}_{22}(\omega)\;
\end{pmatrix}
\end{equation}
with
\begin{align*}
\Sigma^{<}_{11}(\omega) &= -i\Gamma_{N}f_N(\omega-\mu_N)-\cfrac{i\Gamma_S|\omega|}{\sqrt{\omega^2-\Delta^2}} \theta(|\omega|-\Delta)f_S(\omega-\mu_S)\\[8pt]
\Sigma^{<}_{12}(\omega) &= \Sigma^{<}_{21}(\omega)=\cfrac{i\Gamma_{S}\Delta}{\sqrt{\omega^2-\Delta^2}}\theta(|\omega|-\Delta)f_S(\omega-\mu_S)\\[8pt]
\Sigma^{<}_{22}(\omega) &= -i\Gamma_{N}f_N(\omega+\mu_N)-\cfrac{i\Gamma_S|\omega|}{\sqrt{\omega^2-\Delta^2}} \theta(|\omega|-\Delta)f_S(\omega-\mu_S)
\end{align*}
Now, multiplying matrices in Eqn. \eqref{eq:12}, we get the lesser Green's function for electrons on the QD as
\begin{equation}\label{eq:15}
\begin{aligned}
G^{<}_{d,11}(\omega) &=  i\Gamma_{N}f_N(\omega-\mu_N)|G_{d,11}^{r}(\omega)|^2+
i\Gamma_{N}f_N(\omega+\mu_N)|G_{d,12}^{r}(\omega)|^2+\\[6pt]
& \frac{i\Gamma_S|\omega|}{\sqrt{\omega^2-\Delta_2}}\;\theta(|\omega|-\Delta)f_S(\omega-\mu_S)\times\\[6pt]
&\left[|G_{d,11}^{r}(\omega)|^2+|G_{d,12}^{r}(\omega)|^2
-\frac{2\Delta}{|\omega|} Re\left(G_{d,11}^{r}(\omega).G_{d,12}^{a}(\omega)\right)\right]
\end{aligned}
\end{equation}
where $f_{\alpha\in N,S}(\omega\mp\mu_{\alpha})=\left[{exp((\omega\mp\mu_{\alpha})/k_BT_{\alpha})+1}\right]^{-1}$ is the Fermi-Dirac distribution function of reservoirs with temperature $T_{\alpha}$ and chemical potential $\pm\mu_{\alpha}$ (measured from Fermi level $\mu_f=0$).\\ 
After self-consistent calculation of the occupancy and Green's function, the non-linear thermoelectric transport properties can be calculated using the following formulas. Let the system is under the influence of finite voltage biasing $\mu_N-\mu_S=eV$ (say $\mu_N=eV$ and $\mu_S=0$) and/or temperature gradient $T_N-T_S=\theta$ (say $T_N=T+\theta$ and $T_S=T$). Then, the charge current $I_C$ and heat current $J_Q$ from left to right reservoir  across  the  QD can be expressed as\cite{Meir1992,Haug2008,Wang2006,Wang2014,Yamamoto2015}.
\begin{equation}\label{eq:16}
I_C = I_A+I_{QP};
\end{equation}
with
\begin{align*}
&I_A = \frac{2e}{h} \int{\left[f_N(\omega-eV,T+\theta)-f_N(\omega+eV,T+\theta)\right] T_A(\omega)\;d\omega}\\[8pt]
&I_{QP} = \cfrac{2e}{h} \int{\left[f_N(\omega-eV,T+\theta)-f_S(\omega,T)\right] T_{QP}(\omega)\;d\omega}
\end{align*}
and
\begin{equation}\label{eq:17}
J_Q = J_A+J_{QP};
\end{equation}
with
\begin{align*}
&J_A =\cfrac{-4\;eV}{h} \int{\left[f_N(\omega-eV,T+\theta)-f_N(\omega+eV,T+\theta)\right] T_A(\omega)\;d\omega}=-2VI_A\\[8pt]
&J_{QP} = \cfrac{2}{h} \int{(\omega-eV)\left[f_N(\omega-eV,T+\theta)-f_S(\omega,T)\right] T_{QP}(\omega)\;d\omega}
\end{align*}
Where $I_A(I_{QP})$ and $J_A(J_{QP})$ are Andreev (quasiparticle) contribution to charge and heat current respectively.\\
$T_A(\omega)=\Gamma_N^2 |G_{d,12}^r(\omega)|^2$ is the Andreev tunnelling amplitude and $T_{QP}(\omega) =\cfrac{\Gamma_N\Gamma_S|\omega|}{\sqrt{\omega^2-\Delta^2}}\;\theta(|\omega|-\Delta)\times
\left[|G_{d,11}^r(\omega)|^2+|G_{d,21}^r(\omega)|^2-\cfrac{2\Delta}{|\omega|}Re{(G_{d,11}^r(\omega)\;G_{d,12}^a(\omega))}\right]$
is the quasi-particle tunnelling amplitude.\\
In order to use N-QD-S as a heat engine or power generator, the temperature gradient $\theta$ is set larger then zero. Due to this temperature difference electrons move from left reservoir to the right reservoir and thus create a potential difference ($\mu_N-\mu_S = eV_{th}$) due to accumulation of electrons on the right reservoir and positive charge to the left reservoir. The thermovoltage ($V_{th}$) is determined from the condition\cite{Josefsson2018,Josefsson2019}
\begin{equation} \label{eq:18}
I_C(V_{th},\theta)+\cfrac{V_{th}}{R}=0
\end{equation}
in the presence of external serial load resistance $R$. Eqn \eqref{eq:18} is solved numerically to obtain $V_{th}$ and eventually thermopower $S=\cfrac{V_{th}}{\theta}$ and thermal conductance $K = \cfrac{J_Q}{\theta}$. The finite power output $P=-I_CV_{th}=I_C^2R$ generated by the heat engine dissipates across $R$. The thermoelectric efficiency is defined as the  ratio between the generated output power and the nonlinear input heat current i.e $\eta = P/J_Q$.\\
The maximum power output $P_{max}$ is calculated by optimizing $V_{th}$ and $\epsilon_d$ for different values of external load $R$ and the relative efficiency at maximal power output is given by,
\begin{equation}\label{eq:19}
 \left(\cfrac{\eta_{P_{max}}}{\eta_C}\right) = \cfrac{P_{max}}{J_Q}\times{\cfrac{T+\theta}{\theta}}
\end{equation}
where $\eta_C=\cfrac{\theta}{T+\theta}$ is upper bound Carnot efficiency of the heat engine.\\
Further, it is important to note that the Andreev Joule heating, as well as that of quasiparticle contribution to the heat current (energy carried by quasiparticles + Joule heating) in Eq, \eqref{eq:17}, appear only in the nonlinear regime.  Based on the Peltier effect, the total heat flow $J_Q = J_{QP}+J_A$ removes or adds heat to the normal metal, causing the temperature of the normal metal to decrease or increase, and the system can work as a cooler or refrigerator.
\section{Result and discussion}\label{sec:3}
\begin{figure*}[!htb]
\centering
\includegraphics
  [width=0.9\hsize]
  {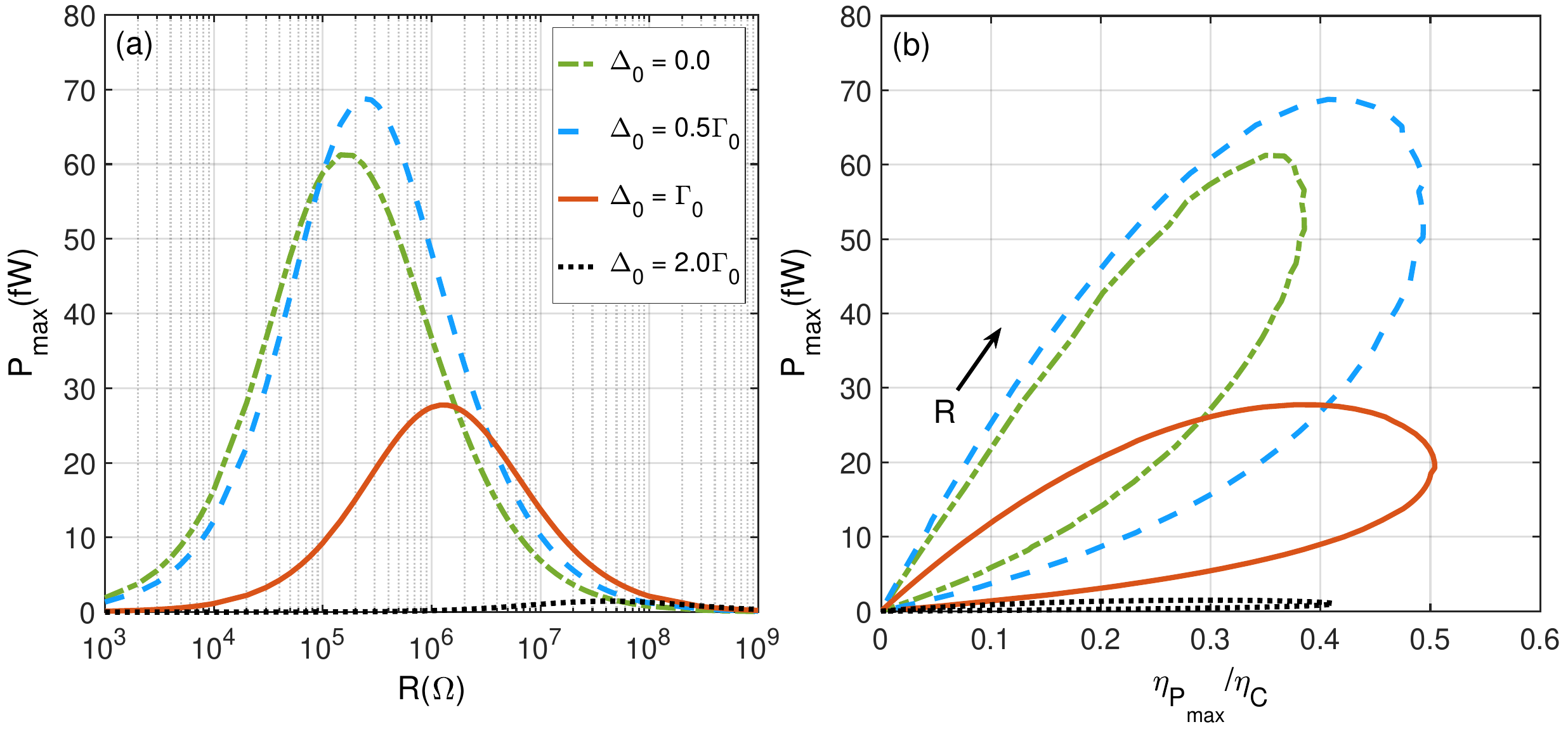}
\includegraphics
  [width=0.9\hsize]
  {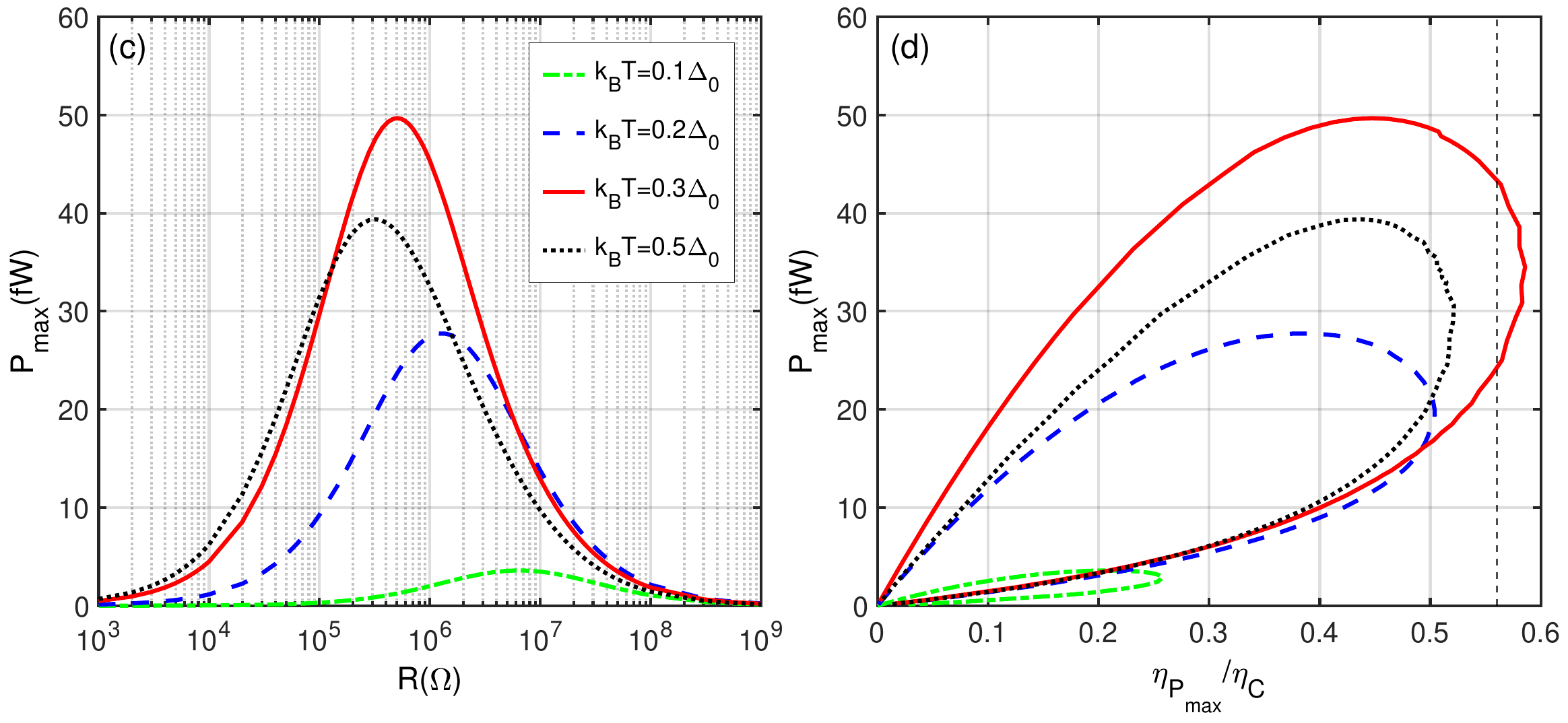}
\caption {(a) $P_{max}$ as a function of external load $R$, and (b) $P_{max}$ as a function of $\eta_{P_{max}}$ for different $\Delta_0$. The other parameters in (a) and (b) are: $U=2.0\Gamma_0$, $\Gamma_S=\Gamma_N=0.1\Gamma_0$, $k_BT=0.2\Gamma_0$, and $k_B\theta=0.1\Gamma_0$. (c) $P_{max}$ as a function of $R$, and (b) $P_{max}$ as a function of $\eta_{P_{max}}$ for different $k_BT$ . The vertical black dashed lines at $\eta_{P_{max}}=0.56\eta_C$ indicate the Curzon–Ahlborn efficiency ($\eta_{CA}$). The other parameters in (c) and (d) are: $\Delta_0=\Gamma_0$, $U=2.0\Gamma_0$, $\Gamma_S=\Gamma_N=0.1\Gamma_0$ and $k_B\theta=0.1\Gamma_0$. The arrow indicates the direction for increasing external load $R$.}
\label{fig:1}
\end{figure*}

\begin{figure*}[!htb]
\centering
\includegraphics
  [width=0.95\hsize]
  {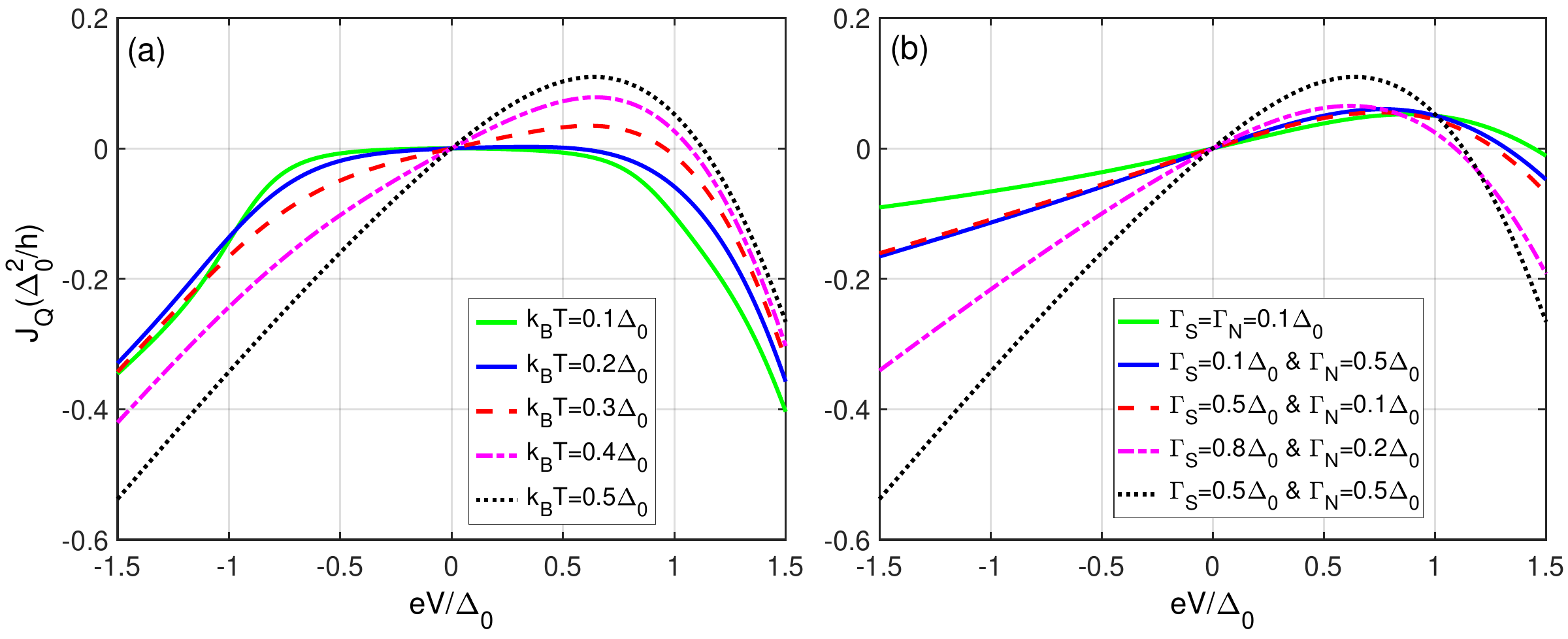}
  \includegraphics
  [width=0.95\hsize]
  {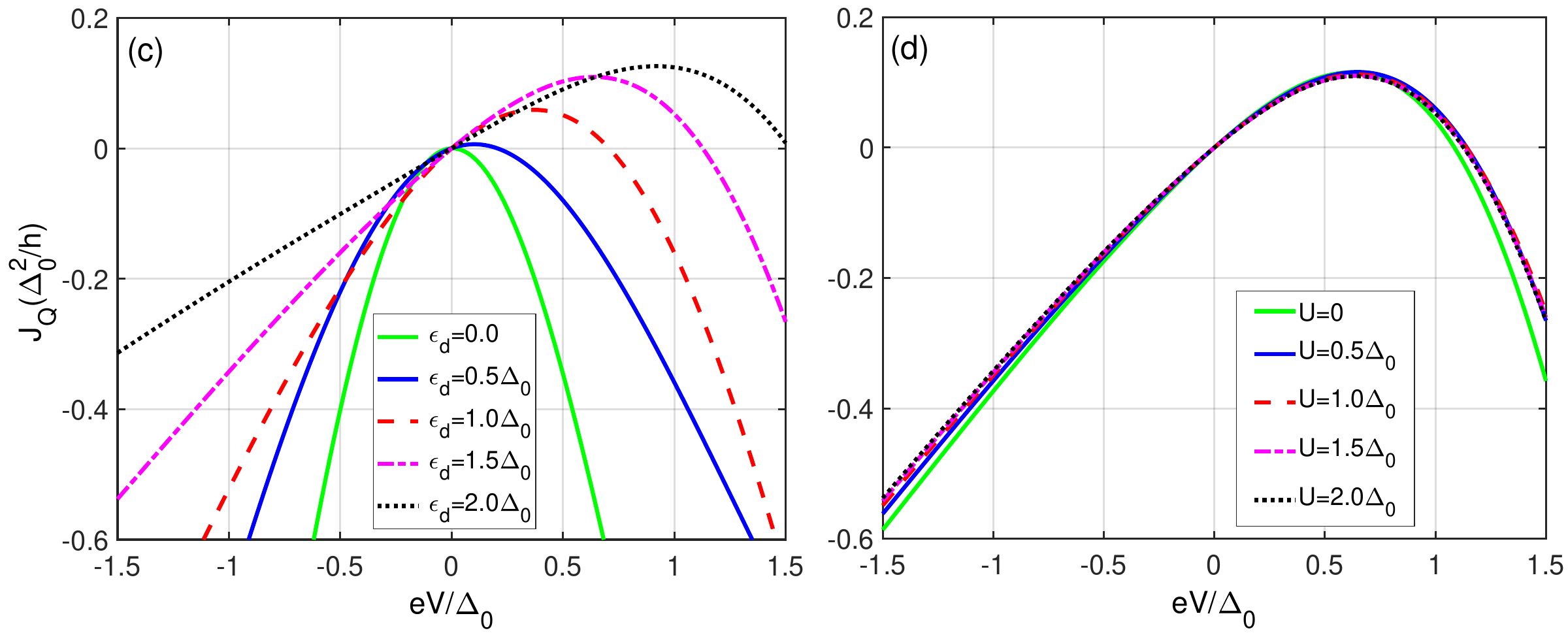}
\caption {Total heat current $J_{Q}$ as a function of bias voltage $eV$ for different (a) background thermal energies $k_BT$ with $\Gamma_S=\Gamma_N=0.5\Delta_0$, $\epsilon_d=1.5\Delta_0$, $U=2\Delta_0$, (b) QD-reservoirs tunneling rates $\Gamma_S/\Gamma_N$ with $k_BT=0.5\Delta_0$, $\epsilon_d=1.5\Delta_0$, $U=2\Delta_0$, (c) QD energy level $\epsilon_d$ with $k_BT=0.5\Delta_0$, $\Gamma_S=\Gamma_N=0.5\Delta_0$, $U=2\Delta_0$, and (b) on-dot Coulomb interaction $U$ with $k_BT=0.5\Delta_0$, $\Gamma_S=\Gamma_N=0.5\Delta_0$, $\epsilon_d=1.5\Delta_0$.}
\label{fig:2}
\end{figure*}

Numerical calculations for the non-linear thermoelectric quantities are done using Matlab based on the equations derived in the previous section and $\Gamma_0$(in $meV$) is considered as the energy unit. We analyze two situations: (1) In Fig. \ref{fig:1}, we discuss the optimal power output and corresponding thermoelectric efficiency of the N–QD–S particle-exchange heat engine, and (2) In Fig. \ref{fig:2}, we consider a voltage-driven case for isothermal reservoirs and discuss the total heat current and Peltier  cooling power as a function of various system parameters.
\\
Fig. \ref{fig:1} shows the variation of maximum power output $P_{max}$ (maximize with respect to the QD energy level $\epsilon_d$ or gate voltage) and normalized efficiency corresponding to the maximum power $\eta_{P_{max}}$ of the NQDS particle-exchange heat engine beyond the linear response regime for different values of superconducting energy gap $\Delta_0$, background thermal energy $k_BT$ and external serial load $R$. It is important to highlight here that Andreev tunneling does not contribute to the creation of thermovoltage ($V_{th}$) and only suppresses it within the superconducting energy gap.  Further, the proximity-induced superconducting gap does not affect the thermoelectric transport properties for the parameter regimes considered in the present work. Therefore, $P_{max}$ and  $\eta_{P_{max}}$ shown here are generated completely by the quasiparticle tunneling close to the superconducting energy gap edge.\\
According to Fig. \ref{fig:1}(a), for a relatively small superconducting gap (say, $\Delta_0=0.5\Gamma_0$), the maximum power output ($P_{max}\approx 70fW$) due to quasparticle tunneling is larger than the $P_{max}$ for NQDN system ($\Delta_0=0$). However, as $\Delta_0$ increases, a large thermal energy or thermal gradient is required for quasiparticle tunneling, and hence the thermovoltage ($V_{th}$) and $P_{max}$ are significantly reduced. For $\Delta_0=2\Gamma_0$, the maximum power output $P_{max}$ becomes of the order of few $fW$. Further, as $\Delta_0$ increases, the optimal load, i.e., $R$ corresponding to the peak in $P_{max}$, shifts towards larger values due to the load matching. The optimal load shifts from approximately $100k\Omega$ to $100M\Omega$ as $\Delta_0$ increases from $0$ to $2\Gamma_0$. Similar to the maximum power output, for a relatively small superconducting gap, the corresponding thermoelectric efficiency for the NQDS system is greater as compared to the NQDN system, with $\eta_{P_{max}}\approx 50\%\eta_C$ as shown in Fig. \ref{fig:1}(b). As $\Delta_0$ is increased from $0.5\Gamma_0$ to $\Gamma_0$, the value of $\eta_{P_{max}}$ remains almost constant and for $\Delta_0=\Gamma_0$, $\eta_{P_{max}}$ reduces by approximately $20\%\eta_C$.\\
As background thermal energy $k_BT$ increases, initially, $P_{max}$ is enhanced due to increasing quasiparticle tunneling from the normal metallic side, as seen in Fig. \ref{fig:1}(c). When $k_BT$ approaches the energy corresponding to the superconducting transition temperature $k_BT_c$, the $P_{max}$ begins to decrease due to backward hole tunneling from the superconducting side. The thermoelectric efficiency $\eta_{P_{max}}$ in Fig \ref{fig:1}(d) follows a similar behavior as that of $P_{max}$ with increasing $k_BT$. The normalized $\eta_{P_{max}}$ can reach upto $58\%\eta_C$ with Power output $\approx 35fW$ for $k_BT=0.3\Gamma_0$ and $k_B\theta=0.1\Gamma_0$.
\\
Fig. \ref{fig:2} shows the variation of non-linear heat current $J_Q=J_A+J_{QP}$  given in Eqn. \eqref{eq:17} as a function of bias voltage(both forward and reverse bias) for different values of background thermal energy $k_BT$, QD-reservoir tunneling strengths $\Gamma_{S}$/$\Gamma_{N}$, QD level energy $\epsilon_d$ and on-dot Coulomb interaction $U$. The Peltier cooling effect occurs when $J_Q>0$, i.e., heat is absorbed from the normal metallic side. If $J_Q<0$, then Joule heating is dominant, and hence, the applied voltage bias only heats the normal metallic reservoir. Further, due to the particle-hole symmetric nature of Andreev bound states, the Andreev heat current $J_A$ does not contain the energy current, unlike the quasiparticle current $J_{QP}$. Therefore, the Andreev bound states only contribute to the Joule heating or heat dissipation, and the cooling effect is entirely due to quasiparticle tunneling.\\
Fig. \ref{fig:2}(a) shows the heat current $J_Q$ as a function of bias voltage $eV$ for different values of the background thermal energy $k_BT$. At low background thermal energies, i.e., $k_BT\leq 0.1\Delta_0$, no cooling effect ($J_Q>0$) is observed as the quasiparticle tunneling is strongly suppressed and Andreev Joule heating is significant. In this case, the heat current only consists of Joule heating ($J_Q<0$) generated by Andreev and quasiparticle currents and has been analyzed in reference\cite{Verma2022}. Now, for higher temperatures, i.e., $k_BT=0.3\Delta_0$ and $k_BT=0.5\Delta_0$ the cooling effect is observed for the forward bias (positive) voltage within the superconducting energy gap, i.e., $eV\lesssim\Delta_0$. The cooling effect is maximum as $k_BT$ approaches the superconducting transition temperature due to vanishing Andreev current or Joule heating. Now, for large voltages ($eV\gtrsim\Delta_0$), the quasiparticle Joule heating predominates, and the cooling effect ceases. The negative bias voltage($eV<0$) only heats the normal metal reservoir without Peltier cooling.\\
In Fig. \ref{fig:2}(b), the heat current $J_Q$ shows the cooling effect when total coupling strength $\Gamma_N+\Gamma_S$ is varied from $0.1\Delta_0$ to $\Delta_0$. The cooling effect for $eV<\Delta_0$ and $\Gamma_N+\Gamma_S\leq 0.6\Delta_0$ is nearly independent of the ratio $\Gamma_S/\Gamma_N$ and attains a relatively large magnitude for the strong symmetric coupling configuration with $\Gamma_N+\Gamma_S=\Delta_0$ as considered in reference\cite{Hwang2023}.\\
Fig. \ref{fig:2}(c) shows the heat current $J_Q$ as a function of bias voltage $eV$ for different values of the QD level position $\epsilon_d$. When $\epsilon_d$ lies within the superconducting gap, i.e., $\epsilon_d<\Delta_0$, the Andreev and quasiparticle current generates large Joule heating effects even at low bias voltages. The Peltier cooling effect caused by the quasiparticles begins to dominate and extends to the larger bias voltage as the position of QD energy level $\epsilon_d$ is tuned far above the Fermi energy level, i.e., $\epsilon_d\gtrsim\Delta_0$. On the other hand, the large thermal energy, i.e., $ k_BT=0.5\Delta_0$, destroys the Coulomb blockade effect. Therefore, the heat current $J_Q$, including the cooling effect and Joule heating current, is independent of the strength of on-dot Coulomb interaction, as shown in Fig. \ref{fig:2}(d).
\section{Conclusion}
\label{sec:4}
In summary, we have presented a theoretical study of the (1) Seebeck effect and (2) Peltier effect in a hybrid NQDS nanodevice based on an interacting quantum dot coupled between a normal metal and Bardeen-Cooper-Schrieffer superconductor.\\
In the first case, we show that the presence of a superconducting energy gap ($\Delta_0$) significantly affects the maximum power output ($P_{max}$) and corresponding thermoelectric efficiency ($\eta_{P_{max}}$). For smaller superconducting energy gap ($\Delta_0\leq 0.5\Gamma_0$), quasiparticle tunneling led to higher $P_{max}$ values compared to the NQDN case ($\Delta_0=0$). However, as $\Delta_0$ increases, the need for a large thermal energy $k_BT$ and thermal gradient $k_B\theta$ reduced both $P_{max}$ and $\eta_{P_{max}}$. Background thermal energy ($k_BT$) also played a crucial role, enhancing $P_{max}$ initially but decreasing it as $k_BT$ approached the energy corresponding to superconducting transition temperature $k_BT_c$.\\
In the second case, we analyzed non-linear heat current($J_Q$) as a function of bias voltage and other system parameters. The Peltier cooling ($J_Q>0$) of the normal source reservoir is observed for positive bias voltage($eV>0$), with maximum cooling occurring as $k_BT$ approaches $k_BT_c$. The coupling strength between the quantum dot and reservoirs influenced the heat current, with symmetric and strong coupling configurations leading to larger cooling magnitudes. Additionally, if the quantum dot energy level ($\epsilon_d$) lies within the superconducting gap, Joule heating effects are significant, even at low bias voltages. However, as $\epsilon_d\gtrsim\Delta_0$, the Peltier cooling effect dominates and extends to higher bias voltages. The Coulomb blockade effect became negligible at high thermal energies, making the heat current independent of on-dot Coulomb interaction strength. These results provide significant insights into the non-linear Seebeck and Peltier effect in NQDS nanodevice, with potential applications for the miniature on-chip power generators and refrigerators in cryogenic nanoelectronics.
\section*{Acknowledgements}
Sachin Verma is presently a research scholar at the department of physics IIT Roorkee and is highly thankful to the Ministry of Education (MoE), India, for providing financial support in the form of a Ph.D. fellowship.

\end{document}